\documentclass{iaus}
\usepackage{graphicx}

\title[Star formation in galaxies] 
{Star formation in galaxies: the role of spiral arms}

\author[Clare L. Dobbs]   
{Clare L. Dobbs$^1$
}

\affiliation{$^1$ School of Physics \& Astronomy, University of Exeter, Stocker Road, Exeter, UK, EX4 4QL\\ email: {\tt dobbs@astro.ex.ac.uk \\[\affilskip]
}}

\pubyear{2014}
\volume{298}  
\pagerange{XX -- YY}
\jname{IAUS\,298 Setting the scence for Gaia and LAMOST}
\editors{S. Feltzing, G. Zhao, N.\,A. Walton \& P.\,A. Whitelock, eds.}
\begin{document}

\maketitle

\begin{abstract}
Studying star formation in spiral arms tells us not only about the evolution of star formation, and molecular clouds, but can also tell us about the nature of spiral structure in galaxies. I will address both these topics using the results of recent simulations and observations. Galactic scale simulations are beginning to examine in detail the evolution of GMCs as they form in spiral arms, and then disperse by stellar feedback or shear. The overall timescale for this process appears comparable to the crossing time of the GMCs, a few Myrs for $10^5$ M$_{\odot}$ clouds, 20 Myr or so for more massive GMCs. Both simulations and observations show that the massive clouds are found in the spiral arms, likely as a result of cloud-cloud collisions. Simulations including stars should also tell us about the stellar age distribution in GMCs, and across spiral arms. 
More generally, recent work on spiral galaxies suggests that the dynamics of gas flows in spiral arms are different in longlived and transient spiral arms, resulting in different age patterns in the stars. Such results could be used to help establish the main driver of spiral structure in the Milky Way (Toomre instabilities, the bar, or nearby companion galaxies) in conjunction with future surveys.

\end{abstract}

\firstsection 
\section{Introduction}
Young stars, and molecular clouds, are predominantly situated in the spiral arms of galaxies. One question is whether spiral arms trigger star formation, or whether they simply `rearrange' young stars, or molecular clouds in the galaxy. \cite[Roberts 1969]{Roberts69} considered the response of gas to stellar spiral arms and showed that the gas experiences a strong shock. The sharp rise in density naturally means that the densities required to produce molecular gas, and gravitational collapse to form stars are attained. Following this, the idea that spiral arms trigger star formation was proposed. However in the 1980's, various observational results queried spiral arm triggering of star formation. \cite[Elmegreen \& Elmegreen 1986]{Elmegreen86} compared the star formation rates in flocculent and grand design galaxies, and found that there was little difference despite the different spiral arms. Vogel 1988 supposed that in M51, a galaxy with a high H$_2$ fraction, the arms merely gather together pre-existing molecular clouds. More recent observations still debate this: \cite[Eden et al.  2012]{Eden13} and \cite[Foyle et al.  2010]{Foyle10} find little difference in the star formation efficiencies in spiral arms and inter-arm regions. However \cite[Seigar \& James  2002]{Seigar02} do find a dependence of star formation rate on the strength of spiral arms.

Theoretically, there are two main proposals for how GMCs, the sites of star formation, form; gravitational instabilities and cloud-cloud collisions. Cloud-cloud collisions are increased in spiral arms (\cite[Casoli \& Combes 1982]{Casoli82}, \cite[Dobbs 2008]{Dobbs08}), whilst the strength of the spiral arms will influence the length and timescales over which gravitational instabilities occur, and indeed if the gas is gravitationally unstable or not (\cite[Elmegreen 1978]{Elmegreen78}, \cite[Elmegreen \& Elmegreen 1983]{Elmegreen83}, \cite[Balbus 1988]{Balbus88}).

Spiral galaxies clearly exhibit many different types of structure, and different numbers and shapes of spiral arms. Generally spiral galaxies can be divided into three types: flocculent spirals with many short spiral arms (e.g. NGC 2841), multi-armed with fewer, long spiral arms (e.g. M101), and grand design galaxies (e.g. M81) with typically two spiral arms. 
The first two likely form from gravitational instabilities in the stars and gas, whereas grand design galaxies are usually associated with bars, or tidal companions which drive spiral arms. In the following, we discuss the evolution and properties of clouds, and star formation rates in these different types of spiral galaxy.

\section{GMC evolution in a grand design spiral}
\begin{figure}
\begin{center}
 \includegraphics[width=6in]{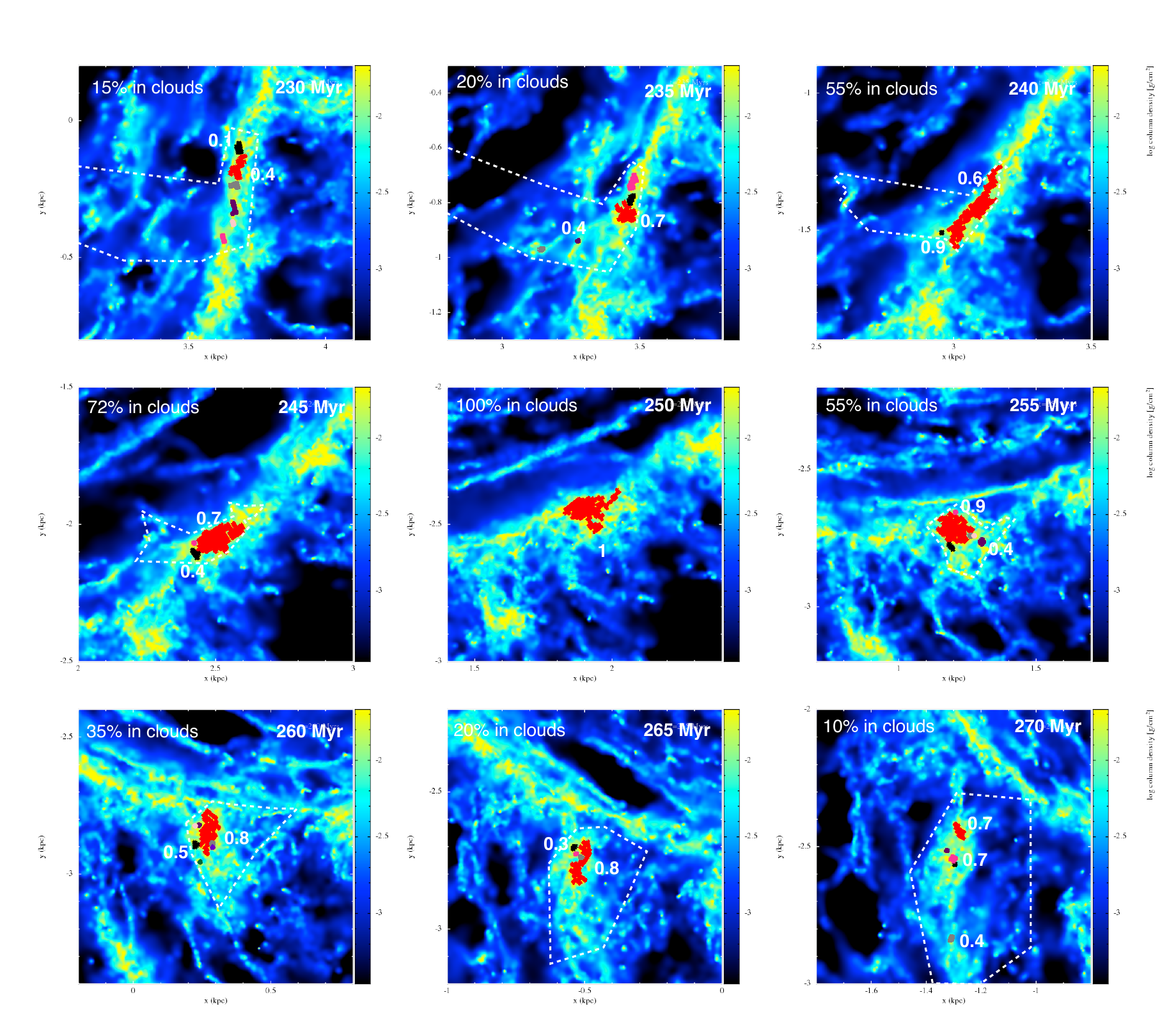} 
 \caption{The evolution of a $2\times10^6$ M$_{\odot}$ GMC is shown over a period of 40 Myr. The GMC was selected at a time of 250 Myr (middle panel) and gas which is contained in the GMC shown at earlier and later times. All the panels are shown in the rotating frame of the spiral potential. Clouds which contain particles of the original GMC are indicated by blocks of colour overplotted on the column density plots. The fraction of particles which are original cloud particles of the chosen GMC are indicated for some of the clouds, for clarity fractions are typically only shown for a couple of clouds. The remainder of the gas which forms / was situated in the original GMC lies in the ambient ISM, the locus of this gas indicated by the dashed white lines in each panel.}
   \label{CLD:fig1} 
\end{center}
\end{figure}

An example of the detailed evolution of a spiral arm GMC, including formation and dispersal, in a galactic context, is shown in \cite[Dobbs \& Pringle 2013]{Dobbs13}. In that paper, we performed a simulation of a grand design spiral, by modelling a gas disc subject to an external potential representing the halo and stars, with an $m=2$ perturbation. The simulation also includes heating and cooling, self gravity and stellar feedback. Star formation in clouds from galaxy simulations has also been followed in \cite[Van Loo et al. 2013]{VanLoo13} (a flocculent galaxy), and \cite[Bonnell et al. 2013]{Bonnell13} (also a grand design with a static potential), but these did not include feedback so the simulations stopped once a large number of stars formed. Figure~\ref{CLD:fig1} shows the evolution of a $2\times10^6$ M$_{\odot}$ cloud, the cloud selected using a clump-finding algorithm at 250 Myr. The evolution of the cloud is complex: the cloud is formed by a number of smaller clouds, as well as diffuse gas, and similarly disperses into other smaller clouds and diffuse gas. Thus the cloud cannot be considered in isolation, rather there are mergers with other smaller clouds, and continuous accretion, and loss, of diffuse gas. A similar cloud tracking device was shown in \cite[Tasker et al. 2009]{Tasker09} for a grid code, there the main interest in the clouds was the merger rate: they found (for a galaxy without any spiral arms) that mergers occurred every 1/5 of an orbit. Here mergers occur much more frequently in the spiral arms, but rarely in the inter arm regions. 

From Figure~\ref{CLD:fig1}, it is evidently difficult to determine a timescale for the lifetime of our example cloud. At 230 and 270 Myr it is difficult to assign an obvious precursor or successor. From 240--260 Myr there is one more prominent cloud though. We estimated cloud lifetimes as the time at which there is a cloud which contains at least half the mass of our selected cloud. Using this definition, our lifetime for the cloud in Figure~\ref{CLD:fig1} is 20 Myr. Typically our lifetimes are quite low, around 4-20 Myr, with lower mass GMCs tending to exhibit shorter lifetimes. Our cloud lifetimes are roughly in agreement with the cloud crossing times, supporting the idea originally put forward by \cite[Elmegreen 2000]{Elmegreen00} that star formation occurs in a crossing time. Over the course of their lifetime, the GMCs in the simulation typically convert a few per cent of their mass into stars.

\section{Differences for simulations with and without spiral arms}
\begin{figure}
\begin{center}
 \includegraphics[width=3.8in, bb=50 0 500 450]{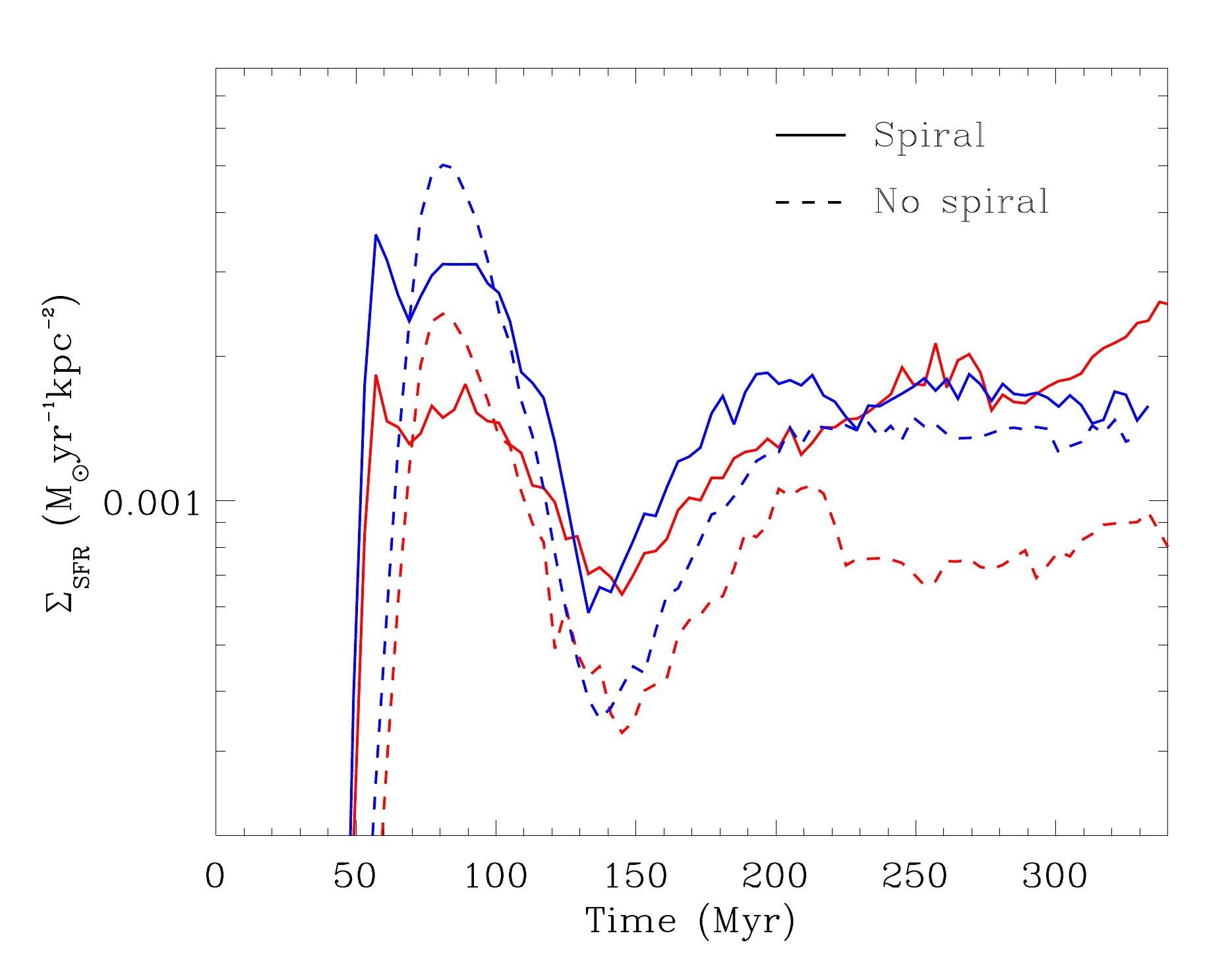} 
 \caption{The star formation rate is shown in simulations with (solid line) and without (dashed line) a spiral potential. The results of simulations with star formation efficiencies of 5 (red, lighter line in black and white version) and 10 (blue, darker line in black and white version) per cent star formation efficiencies are shown. The presence of spiral arms is not found to have a large effect on the star formation rate.}
   \label{CLD:fig2} 
\end{center}
\end{figure}

We have not computed the detailed evolution of clouds in other types of spirals, but in \cite[Dobbs et al. 2011]{Dobbs11}, we compared cloud properties and star formation rates in a galaxy with a spiral potential versus a galaxy with an unperturbed (smooth) stellar disc. In the latter case, the gas is subject to thermal and gravitational instabilities which are sheared out, but there are no real spiral arms.  Figure~\ref{CLD:fig2} shows the star formation rate in simulations with and without spiral arms (with different star formation efficiencies of 5 and 10 \%). There is only a small increase in the star formation rate with spiral arms. So the spiral arms do not appear to trigger star formation. There is a difference in cloud properties though in the cases with and without spiral arms. With spiral arms, we find more massive clouds (see Figure~19 of \cite[Dobbs et al. 2011]{Dobbs11}) of around $10^6$ M$_{\odot}$. However in the absence of spiral arms, there are no clouds $>3\times 10^5$ clouds M$_{\odot}$. Thus the spiral arms appear to gather gas, that would anyway be forming stars, into more massive clouds.

In \cite[Dobbs \& Pringle 2009]{Dobbs09}, we also looked at the amount of bound gas as a measure of star formation in different simulations. There we found that whilst the density increases in stronger shocks, the velocity dispersion also increases, thus there is not a great change in the amount of bound gas with different shock strengths. In the simulations with feedback though, the velocity dispersion is governed by the stellar feedback as well as spiral shocks (see Figure~6 of \cite[Dobbs et al. 2011]{Dobbs11}), so the effect of the spiral shock on the velocity dispersion is more muted. 

In \cite[Dobbs et al. 2012]{Dobbs12}, we also showed differences in the flow of gas, and the constituent gas which forms GMCs in models with and without spiral arms. Without spiral arms, clouds tend to form by gravitational instabilities and mergers between clouds are less important. Consequently gas which forms GMCs tends to be lower density, likely atomic material which undergoes a more sudden phase transition as it is subject to gravitational collapse. For the spiral arm case, gas which forms GMCs tends to be much more of a mixture of clouds as well as more diffuse gas, as seen in Figure~\ref{CLD:fig1}. Furthermore, with spiral arms, there is much clearer evidence for large scale converging flows (corresponding to gas flowing into the spiral arms), as has been supposed as a mechanism for GMC formation (\cite[Heitsch et al. 2006]{Heitsch06}).

The different formation mechanisms for GMCs in the galaxies with and without spiral arms also have some effect on properties of GMCs in addition to those already discussed for the mass of the clouds. Without spiral arms, and fewer cloud-cloud interactions, there tend to be more bound clouds. There are also fewer retrograde rotating clouds (\cite[Dobbs et al. 2011]{Dobbs11}), a phenomenon which arises due to cloud-cloud collisions.

\section{Differences in stellar age distributions for spiral galaxies}
As well as comparing with very flocculent galaxies, we can also compare our grand design, fixed potential example with galaxies where gravitational instabilities predominantly in the stars lead to a small number of relatively long arms. In \cite[Dobbs et al. 2012]{Dobbs12}, we showed an example of such a simulation. In this simulation, there were not such obvious differences in cloud properties compared to the grand design spiral, rather the clouds were relatively similar. Most likely, as there are still relatively strong spiral arms, clouds are still concentrated in the spiral arms, and undergo many interactions there. \cite[Hopkins et al. 2011]{Hopkins11} also performed simulations of these types of multi-armed galaxies, and found similar cloud properties. Any differences in cloud properties are as yet too subtle to be seen clearly in the simulations.
\begin{figure}
\begin{center}
 \includegraphics[width=7in]{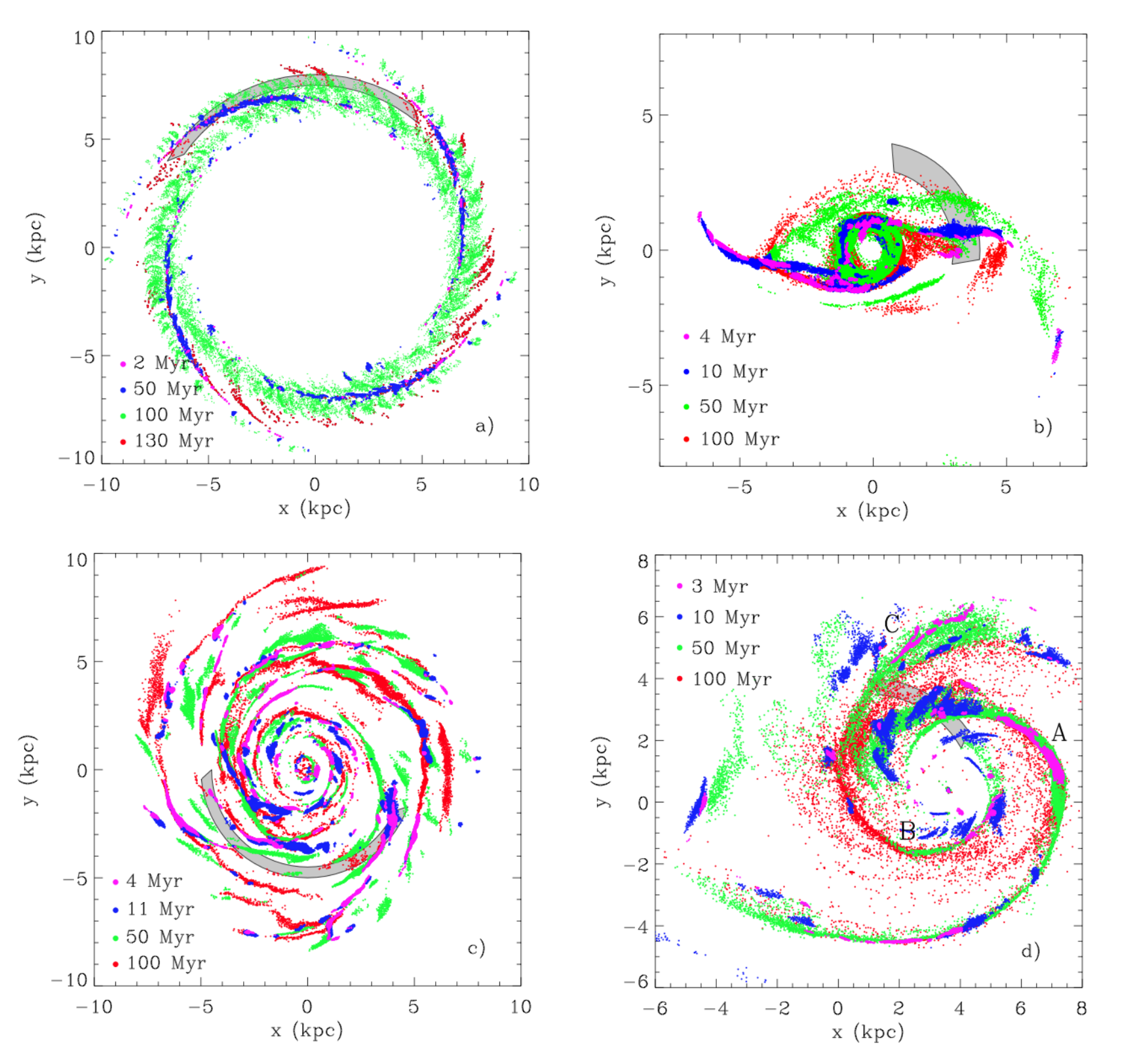} 
 \caption{For galaxy models representing (a) a spiral with a fixed pattern speed, (b) a barred galaxy, (c) a flocculent spiral, and (d) a tidally induced spiral designed to reproduce M51, we show the estimated positions of star clusters of various ages. The galaxies with a relatively constant pattern speed (the fixed spiral (a) and the bar (b)) contain younger stars in the spiral arms or bar, with older (100 Myr) stars downstream in the interarm regions. The distribution of stellar clusters is more complicated in the flocculent (c) and tidally induced spirals (d). For the flocculent galaxy, each segment of a spiral arm tends to contain clusters of a similar age. In contrast, the M51 model shows a complex and somewhat incoherent distribution.}
   \label{CLD:fig3} 
\end{center}
\end{figure}

There are however differences in the gas flow between grand design galaxies, and multi-armed galaxies. In the grand design case, the spiral arms tend to rotate at a quite different rate compared with the rotational velocity of the gas and stars (slower inside corotation). This is likely true regardless of whether the arms are truly quasi-static, slowly winding up with time, or subject to the evolution of the bar. Thus gas travels through the spiral arm. In the multi-armed galaxy, the pattern speed of the arms is much closer to the rotational velocity of the galaxy (\cite[Baba 2013]{Baba13}). Thus gas does not travel through the arm, rather it tends to fall into the local minimum of the potential from both sides of the arm. The gas tends to remain in the arm until the arm disperses. Thus for the two different cases, we would expect quite different stellar age distributions. We investigated this in \cite[Dobbs \& Pringle 2010]{Dobbs10}, where we recorded when gas exceeded a density threshold, and assumed this corresponded to star formation. Thus we were able to make plots of the predicted stellar age distributions (see Figure~\ref{CLD:fig3}). \cite[Grand et al. 2012]{Grand12} have since made plots using actual ages of actual star particles, for a model with a multi-armed barred spiral.

As shown in Figure~\ref{CLD:fig3}, for the grand design galaxy, there is a clear transition of ages moving away from the spiral arms, as would be expected. For the multi-armed galaxy however, there tend to be isolated peaks of stars of the same age. This reflects again that the gas tends to stay in the arms, rather than moving through them. The third panel in Figure~\ref{CLD:fig3} shows a model designed to reproduce M51. In this case, there is also no obvious transition of stellar ages, rather the stellar age distribution is rather chaotic. This is likely because the spiral arms are far removed from quasi-static arms. Instead the arms are quite dynamic, the whole galaxy expanding and contracting with time, as the perturbing galaxy orbits the main M51 galaxy. 

Following Dobbs et al. 10, there have been several observational studies of stellar age distributions in galaxies (\cite[S\'{a}nchez-Gil et al. 2011]{Sanchez11}, \cite[Foyle et al. 2011]{Foyle11}, \cite[Ferreras et al. 2011]{Ferreras 11}). Most of the galaxies examined, including M51, show more chaotic patterns, or stars of similar age along a spiral arm. However there are a couple of examples which show transitions, e.g. M74.

\section{Conclusions}
We have shown the detailed evolution of GMCs for the example of a fixed potential grand design galaxy. The evolution of the GMCs is high complex and non-trivial. Cloud lifetimes are found to be relatively short, and in reasonable agreement with the crossing times of the clouds. We also discussed differences between galaxy simulations with and without spiral arms. With spiral arms, more massive GMCs are formed, but the star formation rates are not significantly different. This supports previous ideas that the main role of the spiral arms is to sweep up material into GMCs rather than directly trigger star formation.

We also discussed using the stellar age distributions of galaxies to test the origin of the spiral arms, i.e. whether they are due to transient gravitational instabilities, or quasi-static density waves, or are somewhat chaotic due to a perturber.



\begin{discussion}

\discuss{Binney}{You said molecular clouds have short lifetimes, $\sim$ 10 Myr, but your conclusion states that the spiral arms merely gather together clouds, which implies they are enduring entities.}

\discuss{Your answer}{Our prescription for determining cloud lifetimes is quite severe, so if a cloud splits in half, we assume it reaches the end of its life (and that any resultant small clouds are different clouds). Also there is no requirement for gas to remain molecular between arms (though this may be true in some galaxies, e.g. M51), rather the gas can lie in HI clouds. Such gas which traverses between the arms in our models is likely to fall below the resolution and density criteria for selecting clouds.}

\end{discussion}

\end{document}